\documentclass[aps,prd,twocolumn,superscriptaddress]{revtex4-2}
\usepackage{graphicx}
\usepackage{float}
\usepackage{amsmath}
\usepackage{amsfonts}
\usepackage{color}
\usepackage{bm}
\usepackage{bbm}
\usepackage{array}
\usepackage{amssymb}
\usepackage{microtype}
\usepackage[utf8]{inputenc}
\usepackage[english]{babel}
\usepackage{soul}
\usepackage{siunitx}
\usepackage{placeins}
\usepackage[colorlinks=true,
    linkcolor=red,
    urlcolor=magenta,
    citecolor=blue,
    anchorcolor=blue]{hyperref}
\usepackage[nameinlink]{cleveref}
\crefname{figure}{Fig.}{Figs.}
\usepackage[caption=false]{subfig}

\newcommand{\mymidrule}{\noalign{\vskip 1.5pt}\hline\noalign{\vskip 1.5pt}}

\begin{document}
	\title{A Surrogate Model for Proton Spectrum Prediction to Map Transitions in Laser-Ion Acceleration}
    \author{Cheng-Qi Zhang}
	\affiliation{
	Key Laboratory of Beam Technology of the Ministry of Education, and School of Physics and Astronomy, Beijing Normal University, Beijing 100875, China}

	\author{Yang He }
	\affiliation{Xinjiang Key Laboratory of Solid State Physics and Devices,
	School of Physics Science and Technology, Xinjiang University, Urumqi 830017, China}
	\author{Mamat Ali Bake }
	\affiliation{Xinjiang Key Laboratory of Solid State Physics and Devices,
	School of Physics Science and Technology, Xinjiang University, Urumqi 830017, China}
    \author{Xilin-Wang }
	\affiliation{School of Systems Science, Beijing Normal University, Beijing 100875, China.}
    
	\author{Bai-Song Xie}\thanks{Corresponding author. Email: bsxie@bnu.edu.cn}
	\affiliation{
	Key Laboratory of Beam Technology of the Ministry of Education, and School of Physics and Astronomy, Beijing Normal University, Beijing 100875, China}
\begin{abstract}
We present a physics-guided, decoupled dual-branch surrogate model to predict continuous proton energy spectra from laser-driven ion acceleration. Integrating a $\beta$-VAE for spectral feature extraction with a parallel multi-layer perceptron for scalar boundary enforcement, the framework achieves a predictive accuracy of $R^2 = 0.94$ for the maximum cutoff energy and $R^2 = 0.94$ for the total particle flux, with a median per-sample spectral $R^2 = 0.985$ (in $\log_{10}$ space) across the full 2000-bin energy distribution. The model incorporates uncertainty quantification via deep ensembles, serving as a quantitative probabilistic diagnostic tool with calibration errors below 6.2\%. Within the 1D longitudinal framework, the surrogate reproduces spectral signatures consistent with the transition from Target Normal Sheath Acceleration (TNSA) to the volumetric heating dynamics of Relativistically Induced Transparency (RIT) and Breakout Afterburner (BOA) regimes, as validated against kinetic diagnostics from 1D particle-in-cell simulations. This approach establishes a computationally efficient baseline for future multi-fidelity optimization and provides an engine for closed-loop parameter control in high-repetition-rate laser facilities.
\end{abstract}
\maketitle

\section{Introduction}\label{sec:intro}
Laser-driven ion acceleration represents a transformative frontier in high-energy-density physics~\cite{Daido2012,Macchi2013IonAB}, driven by the rapid commissioning of petawatt-class laser facilities worldwide~\cite{Danson2019} and the push towards high-repetition-rate applications~\cite{albert20212020}. Since the seminal demonstration of multi-MeV proton beams from solid targets irradiated by ultra-intense laser pulses~\cite{Snavely2000}, a variety of scientific and technological applications have emerged, such as oncological hadron therapy~\cite{Bulanov2002}, fast ignition of inertial confinement fusion~\cite{Roth2001}, and the generation of ultrafast neutron sources~\cite{bayanov1998accelerator}. These applications require precise control over beam energy, spectral shape, as well as sufficient reproducibility and stability~\cite{brack2020spectral}.

Various acceleration regimes have been proposed.  At moderate normalized laser amplitudes, the dominant mechanism is target normal sheath acceleration (TNSA), wherein relativistically heated electrons establish a quasi-electrostatic sheath field at the target rear surface\cite{Wilks1992,passoni2010target}. 
As $a_0$ increases and target thickness $D$ decreases, target thinning or plasma expansion leads to the onset of relativistic induced transparency (RIT), which permits the laser field to penetrate the target bulk, activating volumetric heating and mechanisms such as the breakout afterburner (BOA) \cite{Yin2011,Henig2009}. Achieving precise control and stability for high-repetition-rate proton beams remains a key challenge, requiring the optimization of complex laser-plasma interaction variables\cite{streeter2025stable}.
Besides, the cutoff energy and spectral profile vary sharply with the non-linear coupling of laser and target parameters\cite{schollmeier2015laser,zeil2010scaling,simpson2021scaling}. While particle-in-cell (PIC) simulations \cite{tskhakaya2007particle} provide the standard framework for modeling these interactions, the substantial computational load limits the exploration of full parameter space \cite{kube2021machine}, creating a bottleneck in experimental target design.

To overcome these computational barriers, machine learning (ML) \cite{saravanan2018state} techniques have recently emerged as powerful tools. Initial efforts successfully utilized ML as fast surrogate models to predict scalar beam properties. For instance, Djordjević et al.\cite{djordjevic2021modeling} developed a neural-network surrogate using 1D PIC simulations to model TNSA, uncovering a highly sensitive dependence of maximum ion energy on the pre-plasma gradient length scale. Similarly, Liu et al.\cite{liu2024deep} applied deep learning to an ensemble of 355 1D PIC simulations to predict proton energy in the radiation pressure acceleration regime. Additionally, transfer learning has been employed to predict maximum proton energy by pre-training on 1D data and fine-tuning with sparse 2D simulations\cite{djordjevic2023transfer}.

Concurrently, Bayesian optimization (BO) algorithms\cite{snoek2012practical}have been deployed to actively maximize scalar outputs. While Dolier et al.\cite{dolier2022multi} implemented BO across multi-parameter spaces in simulations, Loughran et al. \cite{loughran2023automated}and Catrix et al.\cite{catrix202520}demonstrated closed-loop BO in high-repetition-rate experiments, achieving unprecedented enhancements in maximum proton energies by tuning target positions and laser wavefronts.

While predicting and optimizing scalar metrics  provides valuable insights into acceleration limits, it is insufficient for translating laser-driven ion sources into practical technologies\cite{Daido2012}. Advanced applications like FLASH radiotherapy \cite{esplen2020physics,romano2022ultra}, radiobiology\cite{friedl2022radiobiology}, and material damage testing rely on precise radiation dose delivery. Achieving this control requires comprehensive knowledge of both particle flux and the full energy distribution.

Unlike predicting a single scalar value, modeling an entire spectral distribution introduces a high-dimensional challenge. A continuous energy spectrum typically comprises thousands of correlated data points, rendering traditional regression techniques computationally intractable and prone to overfitting \cite{goodfellow2016deep}. To address this, the Variational Autoencoder (VAE) was proposed as a generative framework for feature extraction \cite{kingma2013auto}. Rather than performing deterministic compression, the VAE utilizes variational inference to map complex spectral data into a continuous, lower-dimensional probabilistic latent space \cite{bengio2013representation}. Building upon this foundation, the $\beta$-VAE architecture \cite{burgess2018understanding} introduces an adjustable hyperparameter $\beta$ to penalize the Kullback-Leibler (KL) divergence, a statistical metric that measures the deviation of the learned latent distribution from a standard normal prior \cite{kullback1951information}. By constraining this divergence, the model enforces a continuous, structured latent representation that facilitates smooth interpolation across the parameter space, which is required for the continuous reconstruction of physical spectra.
\begin{figure*}[htbp] 
    \centering
    \includegraphics[width=1.05\textwidth]{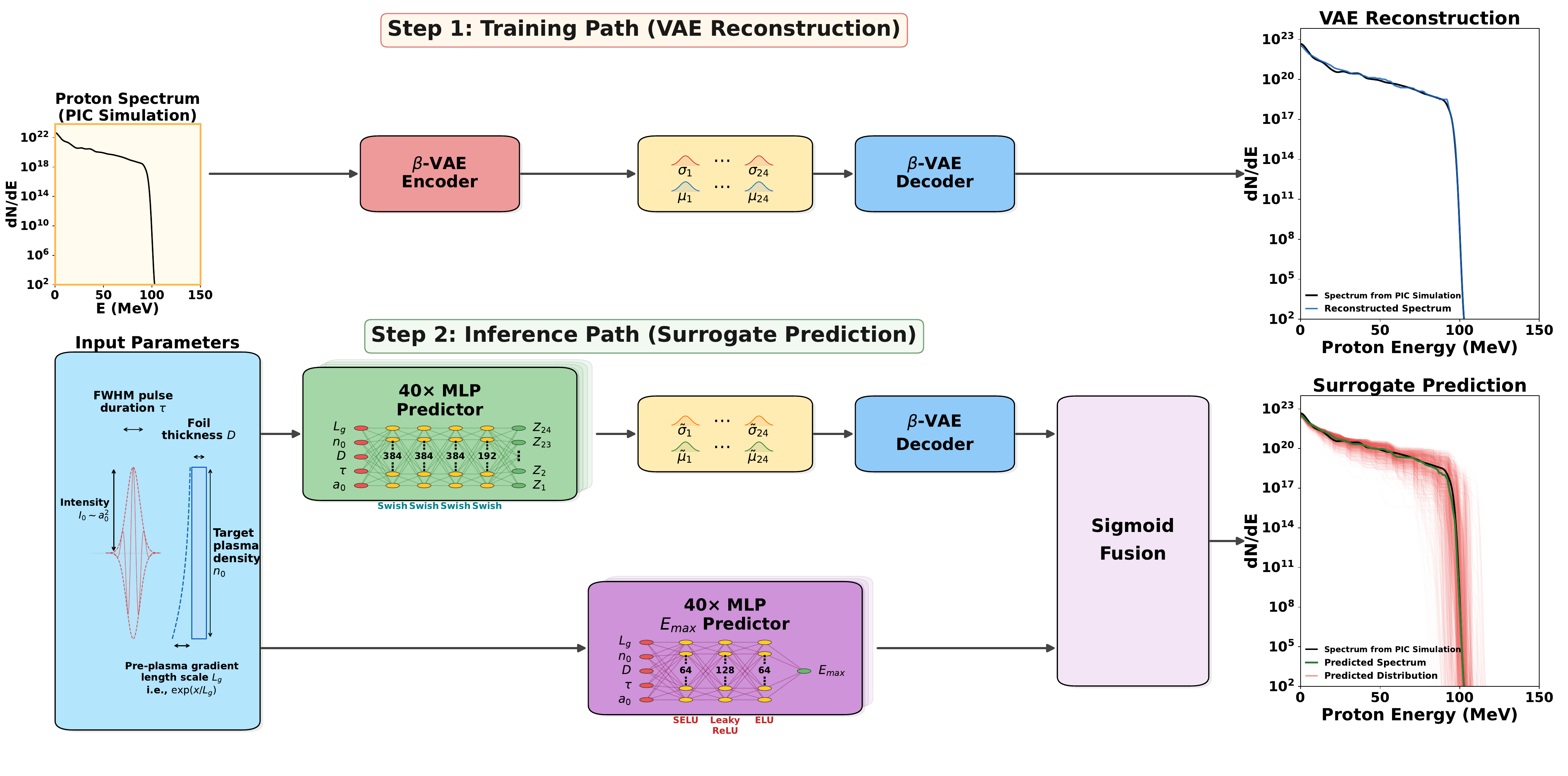}
    \caption{Schematic overview of the decoupled dual-branch surrogate architecture. (Step 1) The training path, where a $\beta$-VAE is optimized to compress and reconstruct the continuous high-dimensional proton energy spectrum; (Step 2) The inference path, where an ensemble of 40 independently trained predictive networks maps initial physical parameters to the latent space, while a parallel MLP branch explicitly predicts the maximum cutoff energy ($E_{max}$). The outputs are finally combined via a Sigmoid fusion mechanism to produce the physically bounded energy spectrum.}
    \label{fig:Fig.1}
\end{figure*}
Recently, these generative approaches have been implemented to reconstruct high-dimensional energy spectra in laser-plasma accelerators for non-destructive synthetic diagnostics. For instance, Streeter et al.\cite{streeter2023laser} utilized VAE to map secondary laser and plasma diagnostics into a low-dimensional latent space, predicting electron energy spectra from Laser Wakefield Accelerators (LWFA). Building on a similar architecture, McQueen et al.\cite{mcqueen2025neural} employed a $\beta$-VAE to compress and predict laser-driven proton spectra based on the statistical moments of experimental back-reflected light. 

\begin{figure*}[htbp] 
    \centering
    \includegraphics[width=1.065\textwidth]{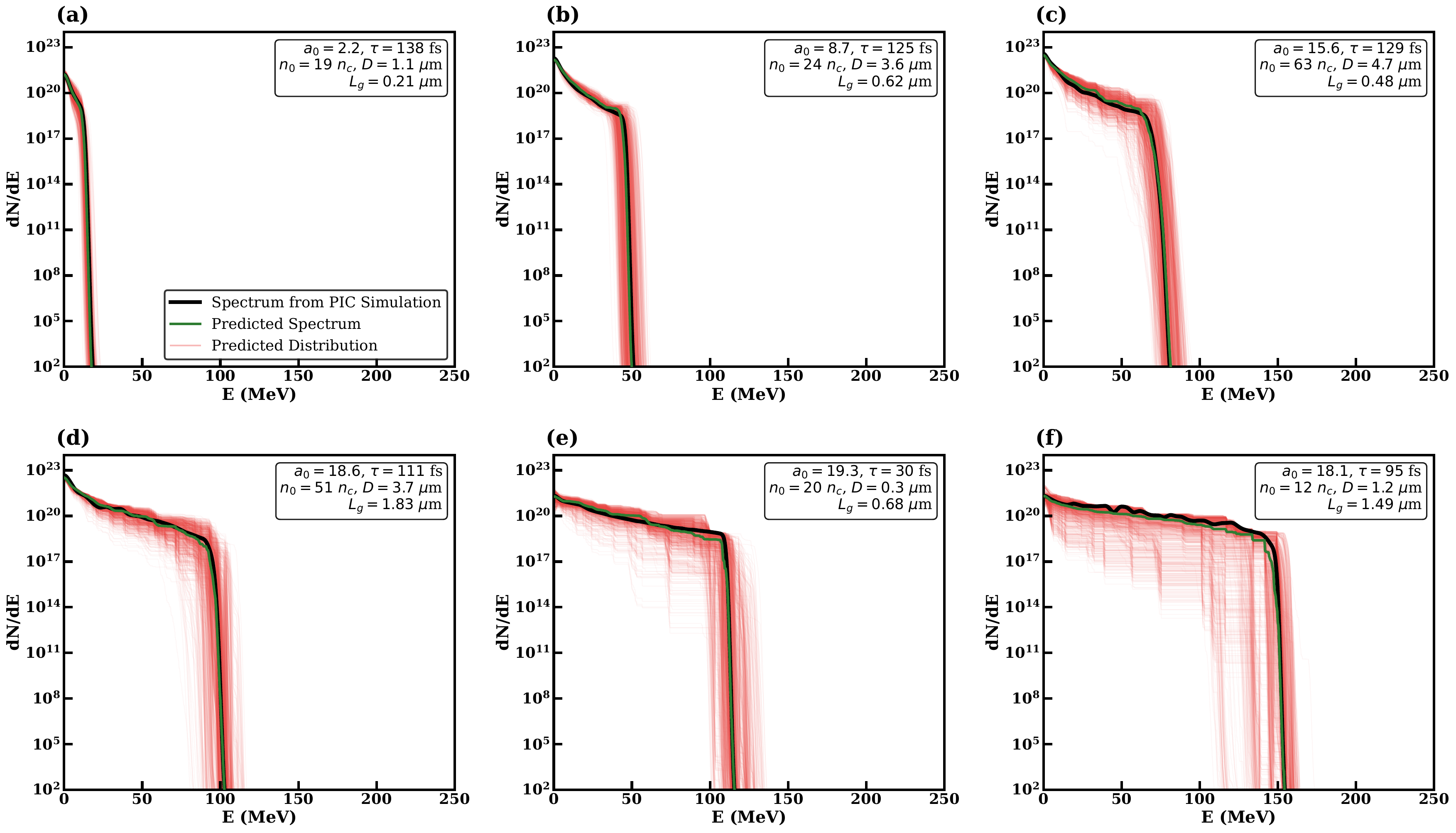}
    \caption{Representative proton energy spectra and prediction accuracy for different parameter settings.}
    \label{fig:Fig.2}
\end{figure*}

Inspired by these generative techniques, our research extends this methodology to predict the full proton energy spectrum directly from initial physical conditions. We developed a hybrid surrogate model that integrates a $\beta$-VAE with a dual-branch multi-layer perceptron (MLP) framework. Specifically, one MLP branch maps the initial conditions to the latent space to capture the overall spectral profile, while an independent MLP explicitly predicts the maximum cutoff energy ($E_{max}$) to enforce a strict physical boundary. Our approach is physics-guided in the sense that analytically derived quantities encoding established plasma scaling laws are injected as input features, and a physics-motivated sigmoid fusion mask enforces spectral termination at the predicted cutoff. Our hybrid surrogate model can predict the full proton energy spectrum from the 5D experimental parameter space, comprising normalized laser amplitude ($a_0$), pulse duration ($\tau$), plasma density ($n_0$), target thickness ($D$), and pre-plasma scale length ($L_g$), to reveal the transition dynamics across diverse acceleration mechanisms.

To verify the generative predictions, we performed extensive 1D-PIC simulations to compare with the model outputs to demonstrate the successful reproduction of the dominant longitudinal dynamics within a 1D framework, where the laser energy coupling and target expansion achieve specific acceleration regimes. Furthermore, our hybrid surrogate model also indicates a marked increase in spectral complexity and prediction uncertainty at extreme relativistic regions ($\gamma > 10$). By exploring scenarios with such extreme laser intensities and lower target densities, we observed the disruption of the quasi-static sheath field due to the onset of relativistic induced transparency, ultimately leading to the generation of complex plateau structures and the termination of the standard TNSA process.

Independent 2D PIC simulations were also conducted as a cross-validation, confirming that the predicted spectral morphology and regime classification are preserved in higher dimensions while the absolute cutoff energies are only modestly reduced. We utilized our physics-guided feature space to investigate the physical reason and found that the drastic variation of the high-energy cutoff occurs when the relativistically modified skin depth surpasses the effective target thickness during the interaction process, causing the termination of opaque acceleration and triggering the highly non-linear BOA process.

\section{Methods}\label{sec:methods}

\subsection{Dataset Generation and Physics-Informed Features}
To construct a high-throughput training dataset, we generated 1,024 one-dimensional PIC simulations spanning a five-dimensional initial parameter space. The input space consists of the normalized laser vector potential ($a_0 \in [2, 20]$), target density ($n_0 \in [5, 80]\,n_c$), pre-plasma scale length ($L_g \in [0.05, 2.0]\,\mu\text{m}$), thickness ($D \in [0.05, 5.0]\,\mu\text{m}$), and pulse duration ($\tau_L \in [30, 150]\,\text{fs}$).
They were drawn using a scrambled Sobol quasi-random sequence to ensure the unbiased and uniform coverage of this multi-dimensional domain \cite{sobol1967distribution,morokoff1994quasi}. Besides, both $a_0$ and $L_g$ were explicitly sampled in logarithmic space to resolve the orders-of-magnitude variations critical to laser intensity and pre-plasma extent.

All simulations were performed using the fully relativistic EPOCH code \cite{arber2015contemporary}. While 1D simulations inherently neglect transverse plasma expansion and multi-dimensional instabilities \cite{palmer2012rayleigh,wan2016physical}, this reduced-dimensionality approach captures the dominant longitudinal electron heating and sheath formation dynamics essential for modeling the overall proton spectral shape. Consequently, it alleviates the computational burden of high-fidelity modeling and serves as an efficient precursor for experimental design. We note that 1D PIC simulations are known to overestimate absolute proton energies relative to multi-dimensional calculations~\cite{djordjevic2023transfer}, because the 1D geometry suppresses transverse dilution of the sheath field. The absolute energy values reported herein should therefore be interpreted within this context; however, the spectral morphologies and regime-transition boundaries remain qualitatively robust. This foundational model also establishes an ideal baseline for future multi-fidelity optimization, where the architecture can be readily fine-tuned using sparse multi-dimensional empirical data \cite{djordjevic2023transfer}. 

The simulation domain spans $[-10, 90]\,\mu\text{m}$ with $n_x = 10{,}000$ cells. A simple laser boundary condition is applied at the left boundary with the linearly polarized laser pulse with the $\lambda_0 = 0.8\,\mu\text{m}$ and a Gaussian temporal intensity envelope, where the full-width at half-maximum (FWHM) is equal to $\tau_L$. 
The target consists of a hydrogen plasma and the pre-plasma follows an exponential density profile $n_e(x) = n_0 \exp[(x - x_\text{front})/L_g]$, extending $4L_g$ ahead of the target front surface, while the main target body maintains a uniform density profile at $n_0$. The spatial domain was resolved with 80 cells per laser wavelength ($\Delta x = 0.01\,\mu\text{m}$), utilizing 500 macroparticles per species. 

To facilitate model training and evaluation, the 1,024 simulations were split into an 80\% training set ($N_\text{train} = 819$) and a 20\% held-out test set ($N_\text{test} = 205$) using a fixed random seed. The final-state proton energy spectra were extracted after 800 fs of physical evolution to ensure the proton bunches have entered a ballistic drift phase and the spectra have converged to their asymptotic terminal states. The extracted continuous spectra were then discretized into 2,000 uniform energy bins spanning 0.1 to 280 MeV.

\begin{table}[H]
\caption{The 12-dimensional physics-guided feature vector. The first five entries are the raw simulation inputs with $n_0$ in units of $n_c$, $D$ and $L_g$ in $\mu$m, and $\tau$ in fs. The remaining seven are analytically derived quantities, both dimensional and dimensionless, that encode known plasma scaling laws.}
\label{tab:features}
\begin{ruledtabular}
\begin{tabular}{lll}
Feature & Expression & Physical role \\
\mymidrule
$a_0$ & --- & Laser amplitude \\
$\tau$ & --- & Pulse duration \\
$n_0$ & --- & Electron density \\
$D$ & --- & Target thickness \\
$L_g$ & --- & Pre-plasma scale length \\
$\mathcal{E}_L$ & $a_0^2\tau$ & Laser energy proxy \\
$\sigma$ & $n_0 D$ & Areal density \\
$\gamma$ & $\sqrt{1+a_0^2/2}$ & Lorentz factor \\
$\xi$ & $a_0/\sigma$ & Transparency parameter \\
$\delta_s$ & $\sqrt{\gamma/n_0}$ & Relativistic skin depth \\
$T_W$ & $0.511(\gamma-1)$~MeV & Wilks temperature~\cite{Wilks1992} \\
$\eta$ & $L_g/\delta_s$ & Coupling ratio \\
\end{tabular}
\end{ruledtabular}
\end{table}

Although the raw 5D parameters define the initial setup, relying on them alone makes it difficult for the neural network to capture highly non-linear laser-plasma dynamics\cite{cranmer2020discovering}. To address this, we expanded the raw inputs into a 12-dimensional physics-guided feature space (\cref{tab:features}), which embeds established scaling laws such as ponderomotive electron heating \cite{Wilks1992} and relativistic transparency \cite{Macchi2013IonAB, Henig2009, yin2011three}. Injecting these physical priors provides an inductive bias that not only accelerates model convergence but also improves predictive reliability when crossing sharp regime boundaries or extrapolating into unseen parameter configurations\cite{karniadakis2021physics, zhu2023reliable}.

\begin{figure}[H]
 \centering
 \includegraphics[keepaspectratio, width=0.48\textwidth]{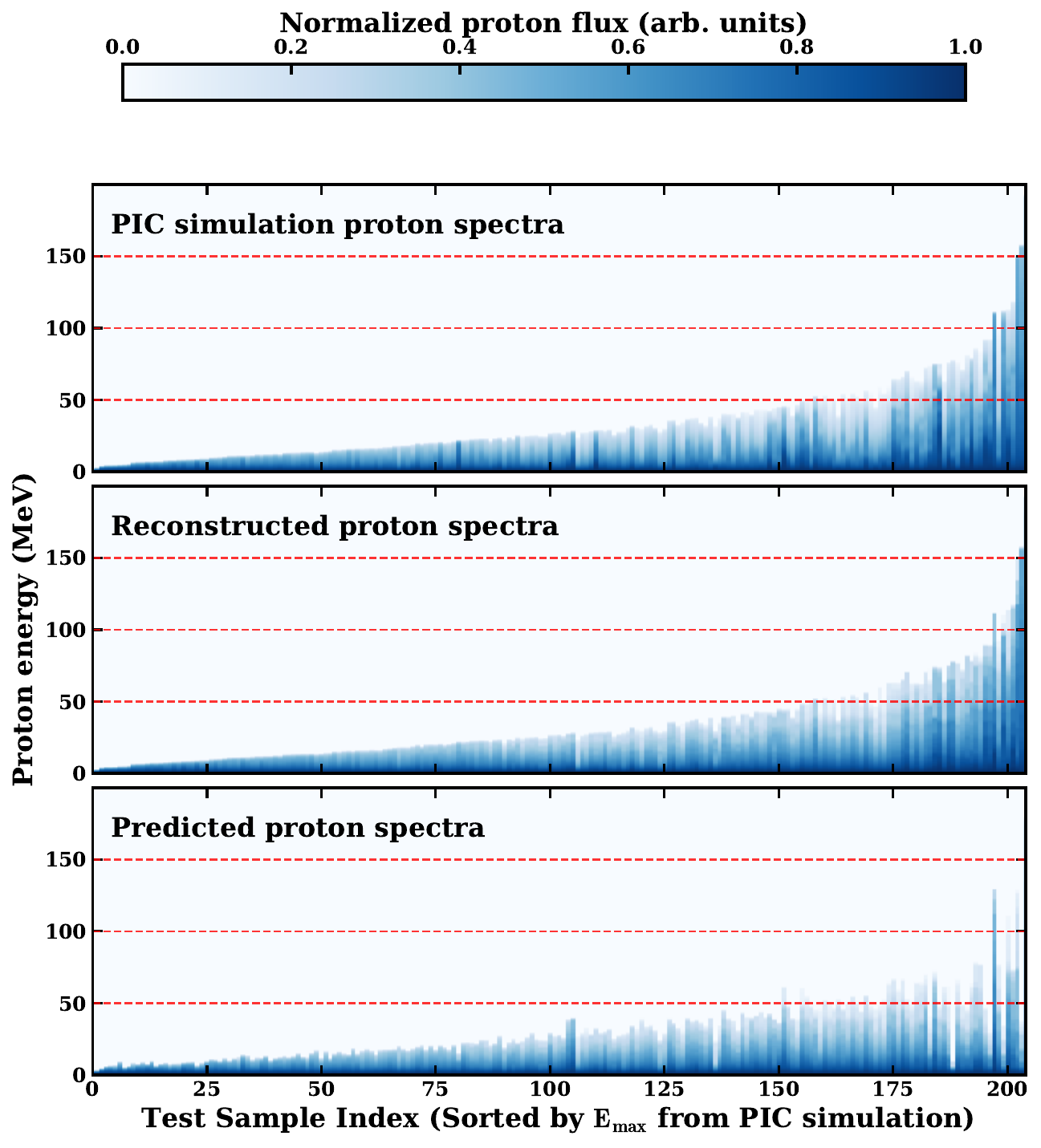}
 \caption{Global visualization and comparison of spectral predictions across the test dataset. The panels display the PIC simulations (top), the $\beta$-VAE reconstructions (middle) and the hybrid surrogate predictions (bottom) respectively sorted by the maximum proton energy from the PIC simulation.}
 \label{fig:3}
\end{figure}

\subsection{Dual-Branch Surrogate Architecture and Training}
The neural network models were implemented in TensorFlow\cite{abadi2016tensorflow}. As noted by Streeter et al.\cite{streeter2023laser} and McQueen et al.\cite{mcqueen2025neural}, global optimization objectives bias such networks toward high-density, low-energy regions, which causes large discrepancies when estimating outliers like the cut-off energy. This issue stems from the continuous latent space smoothing out sharp physical boundaries. To ensure high-fidelity predictions for both the spectral bulk and extreme limits, we propose a decoupled dual-branch surrogate architecture as shown in Fig.\ref{fig:Fig.1}.

The first branch is dedicated to mapping the non-linear morphological evolution of the continuous spectrum. It employs a $\beta$-VAE optimized via an evidence lower bound (ELBO) objective to focus the network on the continuous profile and thermal scaling of the proton bunch. During training, the encoder, configured with ReLU activations \cite{glorot2011deep} and batch normalization, compresses the ground-truth spectra into a 24-dimensional latent space $\mathbf{z}$. The decoder then reconstructs the full spectral profile using ReLU layers and a Sigmoid output activation \cite{han1995influence}. The objective function is designed to enforce latent space structure by integrating three distinct components: a reconstruction loss with adaptive bin weighting, a Kullback-Leibler (KL) divergence penalty ($\beta = 0.001$), and an auxiliary predictor loss ($\beta_\text{aux} = 0.05$) utilizing the non-monotonic Swish activation function \cite{ramachandran2017searching}. The model was optimized using Adam\cite{kingma2014adam} with a batch size of 32 over 1800 epochs, employing a cosine decay learning rate schedule starting at $10^{-3}$. The small $\beta$ value prioritizes reconstruction fidelity over latent-space regularization, a necessary design choice given that the spectral dynamic range spans approximately 20 orders of magnitude in $dN/dE$. While this weak KL penalty does not enforce strong disentanglement in the sense of $\beta$-VAE theory~\cite{burgess2018understanding}, it ensures a continuous, structured latent manifold that supports smooth interpolation across the parameter space.

A separate ensemble of 40 Mixture-of-Experts (MoE)\cite{jacobs1991adaptive} predictor networks is trained to map the 12-dimensional physics-guided input features to this latent space. Each MoE predictor contains two expert sub-networks (utilizing SiLU activations and Dropout) and a lightweight gating network. The gate produces a continuous mixing coefficient $g(\mathbf{x}) \in [0, 1]$ via a sigmoid output, creating a final prediction via soft routing: $\hat{\mathbf{z}} = (1 - g)\mathbf{z}_{\text{expert1}} + g\mathbf{z}_{\text{expert2}}$. The soft routing behaves as a continuous, high-dimensional feature blender. It ensures both experts continuously contribute to every prediction, which accommodates the highly non-linear and continuous nature of the acceleration mechanism transitions observed in our dataset. The MoE ensemble was trained for 1,200 epochs using Adam and a cosine decay learning rate starting at $3 \times 10^{-3}$.
\begin{figure*}[htbp] 
    \centering
    \includegraphics[width=1.05\textwidth]{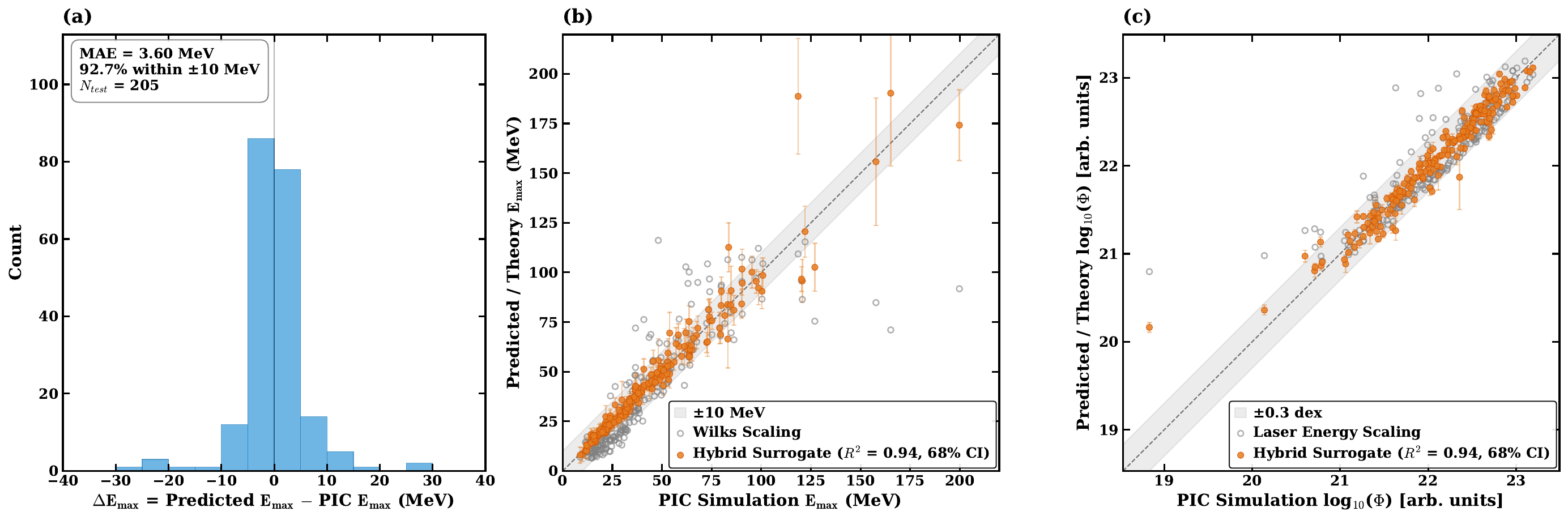}
    \caption{Quantitative evaluation of the  $E_{\max}$ and total particle flux predictions. (a) The error distribution histogram for the predicted $E_{\max}$. (b) Global comparison for the cutoff energy $E_{\max}$. Grey open circles represent the analytical Wilks scaling law, and orange solid dots correspond to the hybrid surrogate predictions. The grey band indicates a $\pm 10$ MeV margin, with error bars representing the 68\% confidence intervals. (c) Comparison for the total particle flux $\Phi$ plotted on a logarithmic scale. Grey open circles denote empirical laser energy scaling, and the grey band represents a $\pm 0.3$ dex margin.}
    \label{fig:Fig.4}
\end{figure*}
In parallel, the second branch deploys an independent deep ensemble of 40 multi-layer perceptron (MLP) networks to predict the absolute $E_{\max}$ boundary. By isolating the scalar cutoff from the high-dimensional generative task, the MLP branch avoids the mathematical smoothing effect characteristic of the VAE. Receiving the same 12-dimensional physics-guided feature set as input, this parallel formulation provides a physics-grounded inductive bias that ensures precise point predictions for the high-energy cutoff. Structurally, the MLP branch utilizes a sequence of SELU\cite{klambauer2017self}, LeakyReLU\cite{maas2013rectifier}, and ELU\cite{clevert2015fast} activations, culminating in a linear output. The MLPs were trained for up to 500 epochs using the Adam optimizer (initialized with a $10^{-3}$ learning rate), with early stopping applied based on a validation patience of 50 epochs. In total, the framework comprises 40 MoE latent-space predictors in the first branch and 40 MLP scalar predictors in the second branch.

The outputs are fused via a differentiable activation mask to recombine the generative and scalar prediction branches. The $\beta$-VAE-reconstructed spectrum, $S_{\text{VAE}}(E)$, is element-wise modulated by an energy-dependent logistic sigmoid function:
\begin{equation}
S_{\text{fused}}(E) = S_{\text{VAE}}(E) \cdot \sigma\left(-\frac{E - E_{\max}^{\text{MLP}}}{w}\right)
\end{equation}
where $E_{\max}^{\mathrm{MLP}}$ is the predicted cutoff energy. The decay width $w$ is sampled from a truncated normal distribution $\mathcal{N}(3.0, 1.5)\,\mathrm{MeV}$, with values clipped to the positive domain ($w > 0$) to prevent non-physical sign reversal. The distribution covers typical physical decay widths across the ensemble members, while the actual regime-dependent spectral morphology is dictated by the $\beta$-VAE. This mathematical formulation enforces a physically consistent and continuous spectral termination. It suppresses flux above the predicted cutoff while preserving the globally learned spectral morphology at lower energies without introducing discontinuities in the gradient flow.

During the training phase, a loss mechanism was implemented to address the severe data imbalance inherent to the physical parameter space, while a target-focal sampling weighting technique \cite{lin2017focal} was applied to the objective function. This prevents the rare, high-value physical states from being overwhelmed by the abundant low-value samples during model optimization. The dynamic sample weight is defined as:
\begin{equation}
s_n = 1 + \alpha \max\left(0, \frac{y_n - \tilde{y}}{y_{\max} - \tilde{y}}\right)
\end{equation}
where $y_n$ represents the target variable of the $n$-th sample, $\tilde{y}$ is the dataset median, and the scaling factor $\alpha$ was set to 1.0 for the VAE and 7.0 for the MoE predictors.

\subsection{Uncertainty Quantification and Recalibration}
To transition our surrogate from a deterministic estimator to a probabilistic diagnostic tool, Uncertainty Quantification (UQ)\cite{lakshminarayanan2017simple} via deep ensembles was established.

An ensemble of 40 models, each with a distinct random initialization but sharing the same training data, was trained concurrently, enabling a rigorous variance decomposition. While this approach forgoes data resampling (all ensemble members see identical training samples, differing only in weight initialization and stochastic training order), the post-hoc recalibration described below corrects any resulting under-dispersion of the raw ensemble spread. By analyzing the predictive scatter among the ensemble members, we decompose the total predictive uncertainty into an aleatoric component (irreducible per-bin residual noise estimated from training-set reconstruction errors) and an epistemic component (inter-model disagreement quantifying knowledge gaps in data-sparse regions). 

To ensure the statistical validity of these confidence bounds, a post-hoc variance recalibration was performed on a held-out calibration subset ($N_\text{cal} = 102$), following the framework of Kuleshov et al.\ \cite{kuleshov2018accurate}. The calibration and evaluation subsets were obtained by a random 50/50 split of the original 205-sample test set using a fixed seed; both subsets were verified to be identically distributed via Kolmogorov--Smirnov tests on all five input parameters ($p > 0.19$ for each). Since the recalibration optimizes only a single scalar parameter $\sigma_{\text{scale}}$ (one degree of freedom), the risk of overfitting to the calibration subset is minimal. A fully independent calibration partition drawn from the training data would further strengthen this guarantee.

The resulting framework achieves root-mean-square calibration errors of 6.1\% for spectra, 2.9\% for $E_{\max}$, and 3.0\% for integrated flux on the independent evaluation set ($N_\text{eval} = 103$). This calibration ensures that the surrogate's predicted confidence intervals represent reliable uncertainty estimates, rendering the model useful as both a prior target optimization tool and as a complement to post-shot synthetic diagnostics.

\section{Results and Discussion}\label{sec:results}
To assess the predictive performance of the decoupled dual-branch surrogate model, we first examine the spectral reconstructions across various physical parameter sets. Fig.\ref{fig:Fig.2} shows representative proton energy spectra for six distinct cases. In each panel the black solid line represents the ground-truth data obtained from PIC simulations, and the red line corresponds to the predictions generated by the hybrid surrogate model. Additionally, the pink shaded region illustrates the 95\% confidence interval derived from the deep ensemble. 

The independent MLP branch allows the model to pinpoint the $E_{\max}$ while the 95\% confidence intervals confirm that the ensemble maintains precision and reliability at the steep boundaries of the spectral distribution.
The surrogate model captures the complex morphological evolution of the particle flux across different interaction conditions.

In Fig.\ref{fig:Fig.2}(a) and Fig.\ref{fig:Fig.2}(b), the spectra exhibit a smooth exponential decay where the relativistically modified skin depth remains smaller than the target thickness. This spectral morphology is typical of surface-heating acceleration regimes such as TNSA. As the laser intensity increases, the standard TNSA process is disrupted. In targets with massive pre-plasmas or extended interactions (panels c, d), hole-boring and dynamic target expansion initiate complex spectral morphologies. Ultimately, when the relativistically modified skin depth approaches the target thickness or the effective density drops below the critical threshold (panels e, f), the interaction enters the RIT regime, triggering volumetric heating and developing high-energy plateaus with abrupt cutoffs.

The robustness of the surrogate model across the entire parameter space is further validated in Fig.\ref{fig:3}, which displays a waterfall visualization of the test dataset. The spectra are sorted by the $E_{\max}$ obtained from PIC simulations to provide a clear comparison of the scaling behavior. Both the $\beta$-VAE-reconstructed and the final predicted spectra show the agreement with the PIC ground truth across the 205 test samples. This consistency underscores the ability of the gating mechanism in MoE networks to dynamically allocate representational capacity across different physical domains.

Beyond qualitative shape generation, precision in predicting scalar metrics is also required. The error distribution histogram in Fig.\ref{fig:Fig.4}(a) shows that for the cutoff energy, the model achieves a Mean Absolute Error (MAE) of 3.60 MeV with 92.7\% of test samples falling within a strict 10 MeV absolute error margin. Fig.\ref{fig:Fig.4}(b) and \ref{fig:Fig.4}(c) provide a comparison between the surrogate model and classical physical scaling laws.

\begin{figure}[H]
\centering
\includegraphics[keepaspectratio, width=0.45\textwidth]{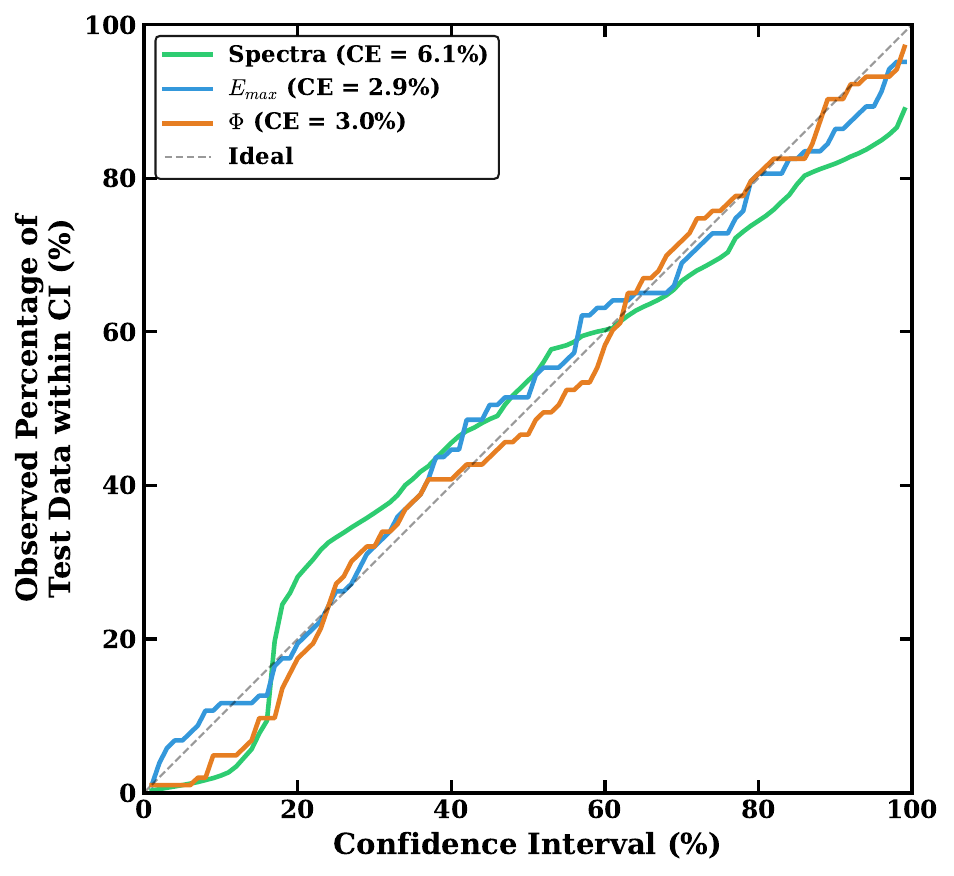}
\caption{Reliability diagrams for the calibration of epistemic uncertainty. The green, blue and orange solid curves represent the calibration performance for the full energy spectra, the cutoff energy $E_{\max}$ and the total particle flux $\Phi$ respectively. The grey dashed diagonal line indicates the ideal case of perfect calibration where the predicted confidence intervals match the observed data frequencies.}
 \label{fig:5}
\end{figure}

\begin{figure}[H]
\centering
\includegraphics[keepaspectratio, width=0.48\textwidth]{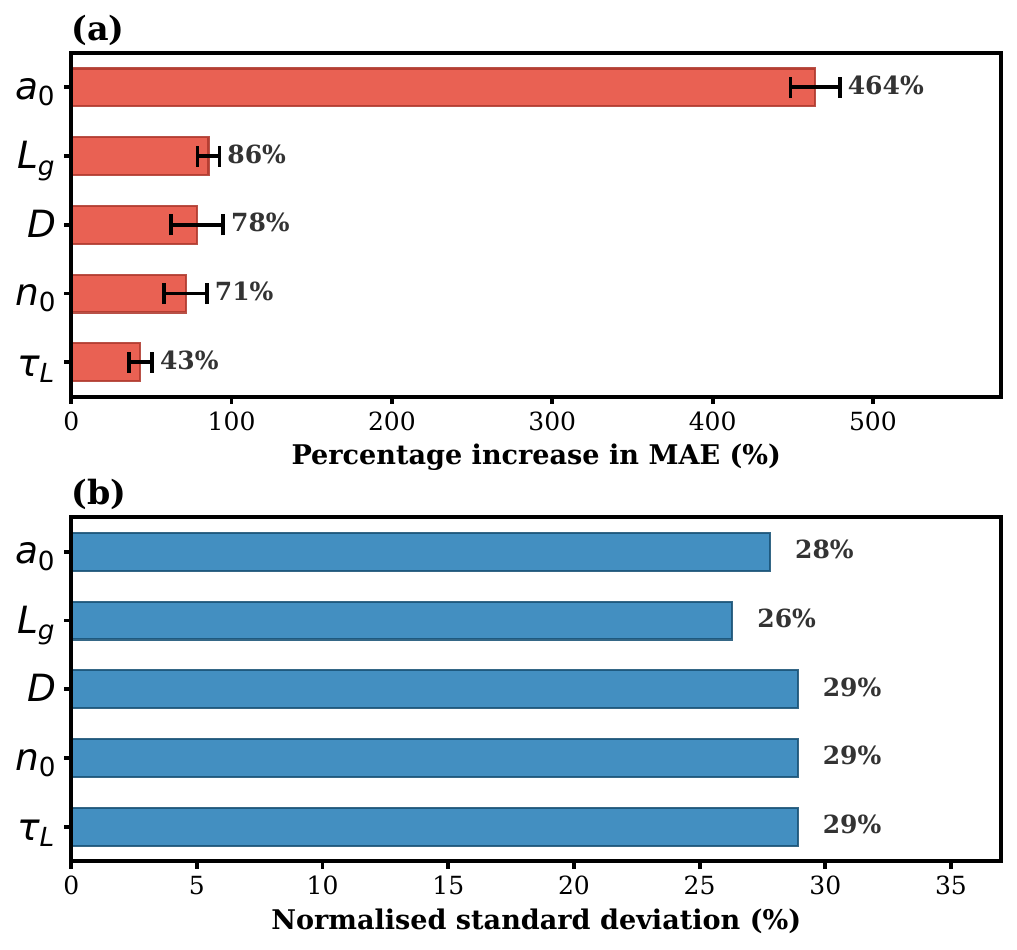}
\caption{Influence of initial physical parameters on surrogate accuracy. (a): Permutation Feature Importance, quantified as the percentage increase in mean absolute error (MAE) when each raw input is independently shuffled across the test set. Shuffling $a_0$ simultaneously disrupts six analytically derived features ($\gamma$, $\xi$, $\mathcal{E}_L$, $\delta_s$, $T_W$, $\eta$), reflecting its role as the dominant upstream driver. (b): normalised standard deviation of each input, confirming that the importance ranking is not an artifact of sampling heterogeneity.}
 \label{fig:6}
\end{figure} 

\begin{figure*}[htbp] 
    \centering
    \includegraphics[width=1.11\textwidth]{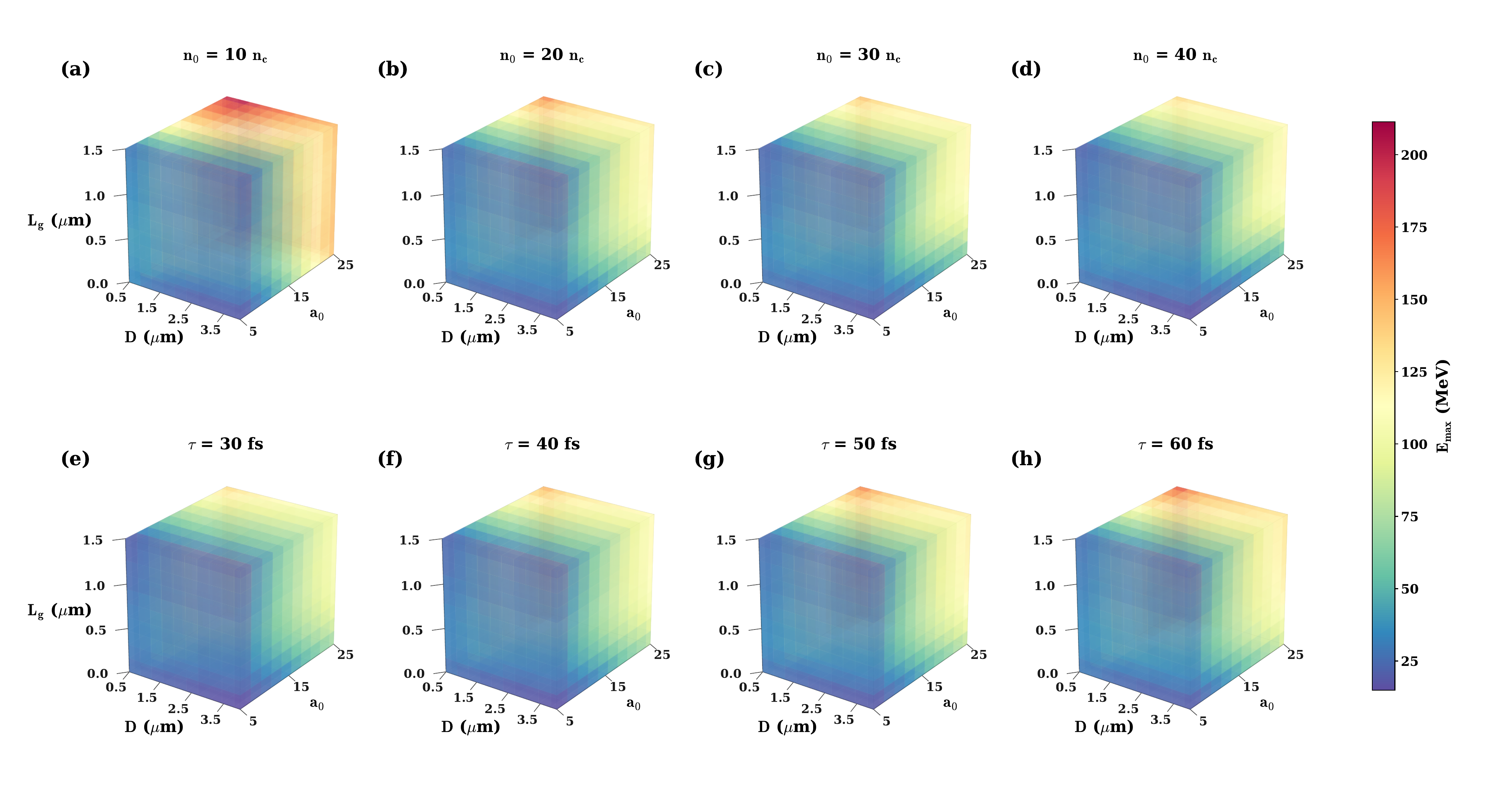}
    \caption{3D data maps of the surrogate-predicted maximum proton energies over the $(D, a_0, L_g)$ parameter cube. Top row (a--d): evaluated at a fixed pulse duration $\tau = 50$ fs for plasma densities $n_0 = 10, 20, 30, \text{ and } 40\,n_c$. Bottom row (e--h): evaluated at a fixed plasma density $n_0 = 20\,n_c$ for pulse durations $\tau = 30, 40, 50, \text{ and } 60$ fs. All panels share a common colour scale.}
    \label{fig:Fig.7}
\end{figure*}

The traditional Wilks scaling law is derived for ponderomotive heating in TNSA regime, which exhibits dispersion when evaluated globally across our dataset. This is inevitable due to the onset of highly non-linear dynamics in extreme relativistic regimes, and relativistic induced transparency \cite{Macchi2013IonAB}.  Our surrogate leverages 12-dimensional physics-guided features to bridge the transitional boundaries between multiple acceleration regimes. The standalone $E_{\max}$ MLP branch achieves a predictive accuracy of $R^2 = 0.94$ (Fig.~\ref{fig:Fig.4}b); after sigmoid fusion with the VAE-reconstructed spectrum, the effective cutoff energy extracted from the fused output retains $R^2 = 0.93$ (Table~\ref{tab:ablation}).Similarly, for the total particle flux $\Phi$, the surrogate achieves an $R^2$ of 0.94, effectively mapping the multi-dimensional parameter space. The $R^2$ for $E_{\max}$ is evaluated in linear space, whereas $R^2$ for $\Phi$ is computed in $\log_{10}$ space to account for its four-order-of-magnitude dynamic range. In addition, the median earth mover's distance (EMD, also known as the Wasserstein-1 distance)\cite{rubner2000earth} between predicted and ground-truth spectra is 0.24~MeV, confirming that the model captures both the spectral shape and absolute normalization to high precision. 

To quantify the contribution of each architectural component and to benchmark the surrogate against conventional approaches, we performed systematic ablation studies and baseline comparisons, summarized in Table~\ref{tab:ablation}. Removing the dedicated $E_{\max}$ MLP branch and relying solely on the $\beta$-VAE output without sigmoid fusion degrades the cutoff-energy coefficient of determination from $R^2 = 0.93$ to $0.87$ and increases the median Wasserstein distance from 0.24 to 0.41~MeV, confirming that the decoupled scalar branch is essential for accurate spectral termination. Furthermore, reducing the 12-dimensional physics-guided input to the raw 5D simulation parameters ($a_0, \tau, n_0, D, L_g$) causes the $E_{\max}$ $R^2$ to drop from 0.93 to 0.80, demonstrating the benefit of injecting analytical scaling-law priors as input features. 

As an external baseline, a direct MLP that regresses the full 2000-bin spectrum without the VAE latent space or the dedicated $E_{\max}$
branch achieves a seemingly comparable per-sample spectral
$R^2 = 0.981$. This metric, however, is dominated by the low-energy bulk and the quiescent region above the cutoff, which together constitute the majority of the 2000 energy bins and are straightforward to fit. The
physically critical scalar metrics expose the true deficiency. The $E_{\max}$ $R^2$ falls to 0.64 and the median Wasserstein distance increases by an order of magnitude to 2.73~MeV, indicating a failure to
localize the spectral cutoff. This result illustrates the bias of global regression objectives toward high-flux, low-energy bins and confirms
that explicit boundary enforcement via the independent $E_{\max}$ branch is the single most impactful component of the proposed framework.

\begin{table*}[htbp]
\caption{Ablation studies and baseline comparison evaluated on the held-out test set ($N_{\text{test}}=205$). The spectral $R^2$ is the median per-sample coefficient of determination computed over all 2000 energy bins in $\log_{10}$ space. The $E_{\max}$ $R^2$ is the coefficient of determination of the cutoff energy extracted from the fused output spectrum in linear MeV space (the standalone $E_{\max}$ MLP branch yields $R^2 = 0.94$ as shown in Fig.~\ref{fig:Fig.4}b; the slightly lower value here reflects the interaction with the sigmoid fusion mask). The $\Phi$ $R^2$ is the coefficient of determination of the total integrated particle flux in $\log_{10}$ space. The EMD is the median earth mover's distance (Wasserstein-1) between the predicted and ground-truth spectral distributions.}
\label{tab:ablation}
\begin{ruledtabular}
\begin{tabular}{l@{\hskip 24pt}c@{\hskip 24pt}c@{\hskip 24pt}c@{\hskip 24pt}c}
Configuration
  & Spectral $R^2$
  & $E_{\max}$ $R^2$
  & $\Phi$ $R^2$
  & EMD (MeV) \\
\mymidrule
Full hybrid (this work)
  & 0.985 & 0.926 & 0.941 & 0.24 \\
Without $E_{\max}$ branch
  & 0.970 & 0.874 & 0.936 & 0.41 \\
5D raw inputs only
  & ---   & 0.799 & 0.934 & ---  \\
Direct MLP regression
  & 0.981 & 0.638 & 0.452 & 2.73 \\
\end{tabular}
\end{ruledtabular}
\end{table*}

Estimating the boundary of its own predictive limits is as crucial as the prediction itself. Fig.\ref{fig:5} presents the reliability diagrams (calibration curves), evaluating whether the predicted confidence intervals represent true statistical probabilities. Through post-hoc variance recalibration, our model achieves alignment, which yields low Calibration Errors (CE=6.1\%) for the full spectra, 2.9\% for $E_{\max}$, and 3.0\% for $\Phi$. These UQ metrics indicate that the deep ensemble is neither overconfident nor artificially conservative. The assigned uncertainty bounds are trustworthy, empowering the surrogate to flag out-of-distribution physical conditions during future experimental campaigns. To empirically verify this capability, we partitioned the test set by laser intensity: samples in the high-$a_0$ tail ($a_0 > 10.9$, top quartile, $N=51$) exhibit a mean epistemic uncertainty $3.2\times$ larger than the in-domain subset ($a_0 \leq 10.9$, $N=154$), confirming that the ensemble spread responds appropriately to data-sparse regions of the parameter space. We note that this high-$a_0$ tail is a sparsely sampled in-domain region rather than a strictly out-of-distribution input; it nonetheless serves as a proxy demonstrating that the ensemble spread grows where training data are scarce, which is the behavior required to flag unreliable predictions in future campaigns.

While the surrogate model achieves a high overall accuracy across the dataset, the predictive precision decreases in the high-energy tails. Consistent with the recent observations by McQueen et al.\cite{mcqueen2025neural}, predicting high-energy spectra presents challenges due to both physical and algorithmic factors. Physically, the relativistic volumetric heating and phase-space dynamics in these regimes introduce strong non-linearities. The resulting spectral morphology is sensitive to small variations in initial parameters, which naturally increases the aleatoric uncertainty. Algorithmically, these conditions represent a sparse region of the parameter space, constituting approximately $\sim 4\%$ of the dataset. Although target-focal sampling mitigates the training bias toward the dominant TNSA regime, the sparsity of the high-energy data still contributes to epistemic uncertainty. Additionally, generative models such as the $\beta$-VAE tend to produce smooth latent representations, making it difficult to fully resolve the sharp spectral cutoffs typical of the BOA mechanism. Consequently, the implemented uncertainty quantification framework provides essential diagnostic value here, as it reflects this expected performance degradation by yielding wider, physically consistent confidence intervals in these non-linear parameter regimes.

The fidelity of the surrogate model in capturing the underlying physical input-output relationships is verified by evaluating the Permutation Feature Importance (PFI)\cite{breiman2001random,fisher2019all}, as presented in Fig.~\ref{fig:6}. This metric quantifies the relative dependency of the model on each raw simulation input by measuring the percentage increase in mean absolute error (MAE, evaluated in $\log_{10}$ space) when that input is independently shuffled across the test set. Because the seven derived features are recomputed from the permuted raw inputs, the importance of each parameter reflects its total upstream influence on the prediction. The results show that the normalized laser amplitude $a_0$ dominates (464\% increase in MAE), consistent with established ponderomotive scaling laws, wherein the absolute kinetic energy limits are fundamentally dictated by the incident laser intensity~\cite{Wilks1992,Macchi2013IonAB}. This finding aligns with the analytical TNSA scaling model of Fuchs et al.~\cite{fuchs2006laser}, in which the hot electron temperature---and hence the maximum proton energy---scales primarily with the laser intensity parameter $a_0$, while geometric factors enter as secondary corrections. A similar dominance of laser intensity over other input parameters was reported by Djordjevi\'{c} et al.~\cite{djordjevic2021modeling}, whose neural-network surrogate trained on over 1{,}000 1D PIC simulations identified a highly sensitive dependence of ion energy on intensity and pre-plasma gradient, with pulse duration playing a comparatively minor role. The remaining parameters---pre-plasma scale length $L_g$ (86\%), target thickness $D$ (78\%), plasma density $n_0$ (71\%), and pulse duration $\tau$ (43\%)---each produce substantial degradation when shuffled, confirming that the surrogate has learned a genuinely multi-dimensional mapping rather than a single-variable proxy. Notably, the pulse duration $\tau$ ranks lowest among all input variables. This is physically consistent with the isothermal expansion models~\cite{fuchs2006laser}, in which $E_{\max}$ depends on $\tau$ only logarithmically through the normalized acceleration time, making $\tau$ inherently a weaker lever than intensity or target geometry. Experimental scaling studies in the sub-picosecond TNSA regime have likewise reported a comparatively weak dependence of $E_{\max}$ on pulse duration at fixed laser energy~\cite{zeil2010scaling}, corroborating the low PFI ranking observed here.

We performed continuous virtual parameter scans of the three-dimensional parameter cube ($D, a_0, L_g$) at different fixed plasma densities ($n_0$). This allowed us to visualize the dynamical processes otherwise hidden by statistical suppression. As shown in Fig.\ref{fig:Fig.7}(a--d), securing maximum proton energies demands high laser intensities and the target geometry must match the plasma density to maintain optimal coupling. The high-energy manifold is localized at lower target thicknesses coupled with extended pre-plasmas to prevent premature target disassembly. As the plasma density escalates, the increased plasma opacity restricts laser penetration. At that identical parameter coordinate, the optimal coupling efficiency degrades, causing the overall maximum energy yield to drop. To verify the role of pulse duration, Fig.\ref{fig:Fig.7}(e--h) presents the same scan at fixed $n_0 = 20\,n_c$ while varying $\tau$ from 30 to 60 fs. The spatial topology of the high-energy manifold remains qualitatively unchanged: increasing $\tau$ produces a systematic but modest upward shift in $E_{\max}$ ($\sim$19\%), consistent with its lowest-ranked permutation feature importance (Fig.\ref{fig:6}). This confirms that the three-dimensional energy landscape is governed primarily by the geometric and intensity parameters ($D, a_0, L_g$), while $\tau$ acts as a secondary scaling factor. These visualization results demonstrate that the cutoff energy is governed by multidimensional interactions. They also reveal the advantage of the surrogate model in instantaneously mapping continuous, high-resolution, multidimensional data spaces that traditional grid-based PIC simulations cannot practically cover.

\begin{figure}[H]
\centering
\includegraphics[keepaspectratio, width=0.45\textwidth]{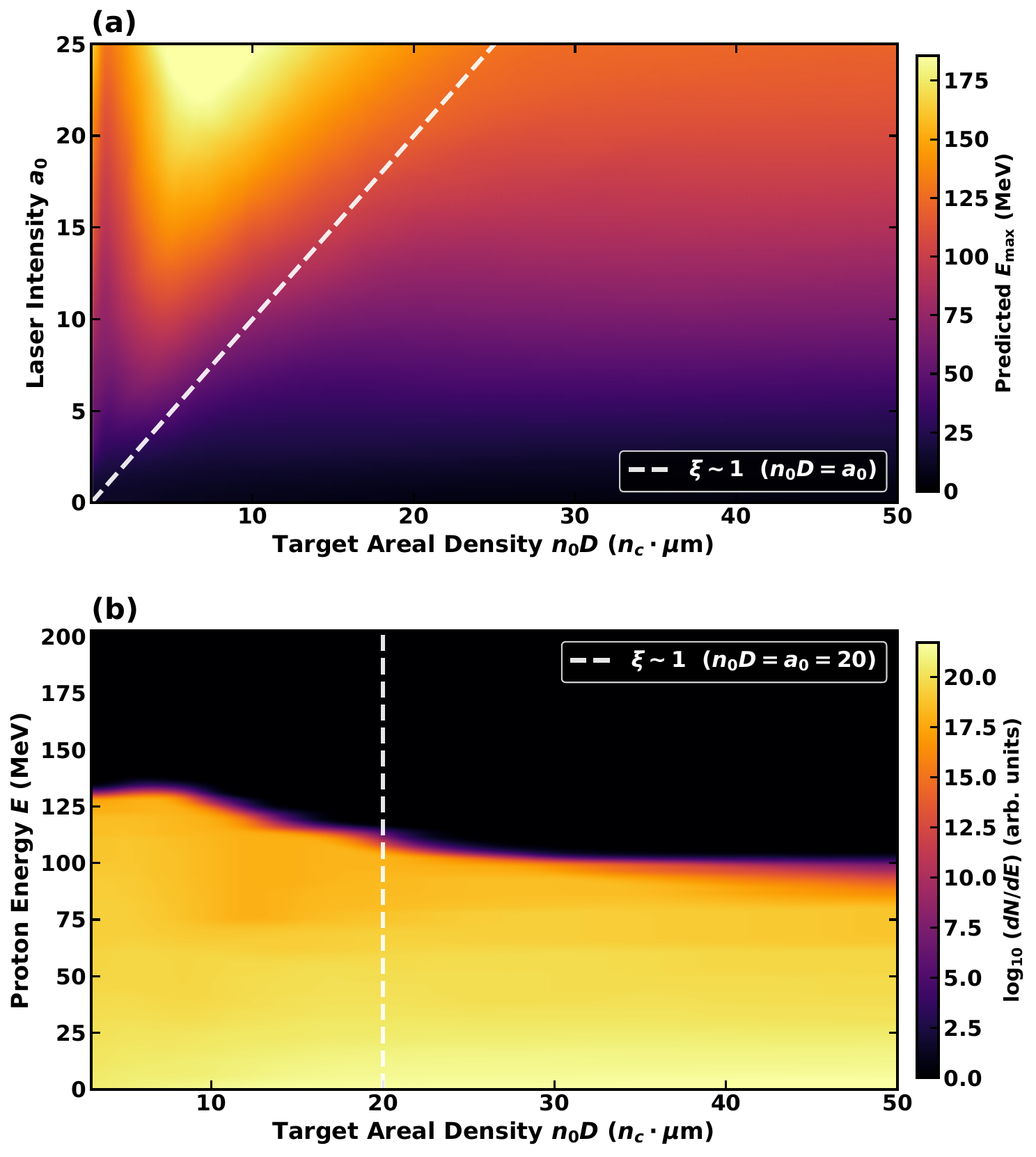}
\caption{Surrogate-mapped acceleration phase diagram and continuous spectral evolution. (a)The 2D phase diagram of the predicted maximum proton energy ($E_{\max}$) across the laser intensity ($a_0$) and target areal density ($n_0D$) domain. (b)A continuous 1D cross-sectional spectrogram tracking the full energy spectrum evolution at a fixed intensity of $a_0 = 20$.}
 \label{fig:8}
\end{figure} 

The surrogate-predicted $E_{\max}$ onto a macroscopic 2D phase diagram of laser intensity ($a_0$) versus this areal density ($n_0D$) is shown in Fig.\ref{fig:8}(a), illustrating the model's ability to reproduce transitions between distinct ion acceleration mechanisms within the 1D framework. In this representation, the laser-target coupling efficiency is governed by the transparency parameter ($\xi = a_0/(n_0D)$). Although the transparency parameter $\xi$ is included as one of the 12 physics-guided input features, the surrogate learns the quantitative location of the RIT threshold ($\xi \sim 1$) as a critical boundary purely from the training data, where the optimal high-energy region (bright yellow) resides predominantly within the relativistically transparent regime.
Fig.\ref{fig:8}(b) provides a continuous 1D cross-sectional slice at a fixed laser intensity of $a_0=20$ to further elucidate this transition, where the previously opaque exponential thermal decay transitions into an extended high-energy plateau. 

\begin{figure*}[htbp] 
    \centering
    \includegraphics[width=1.05\textwidth]{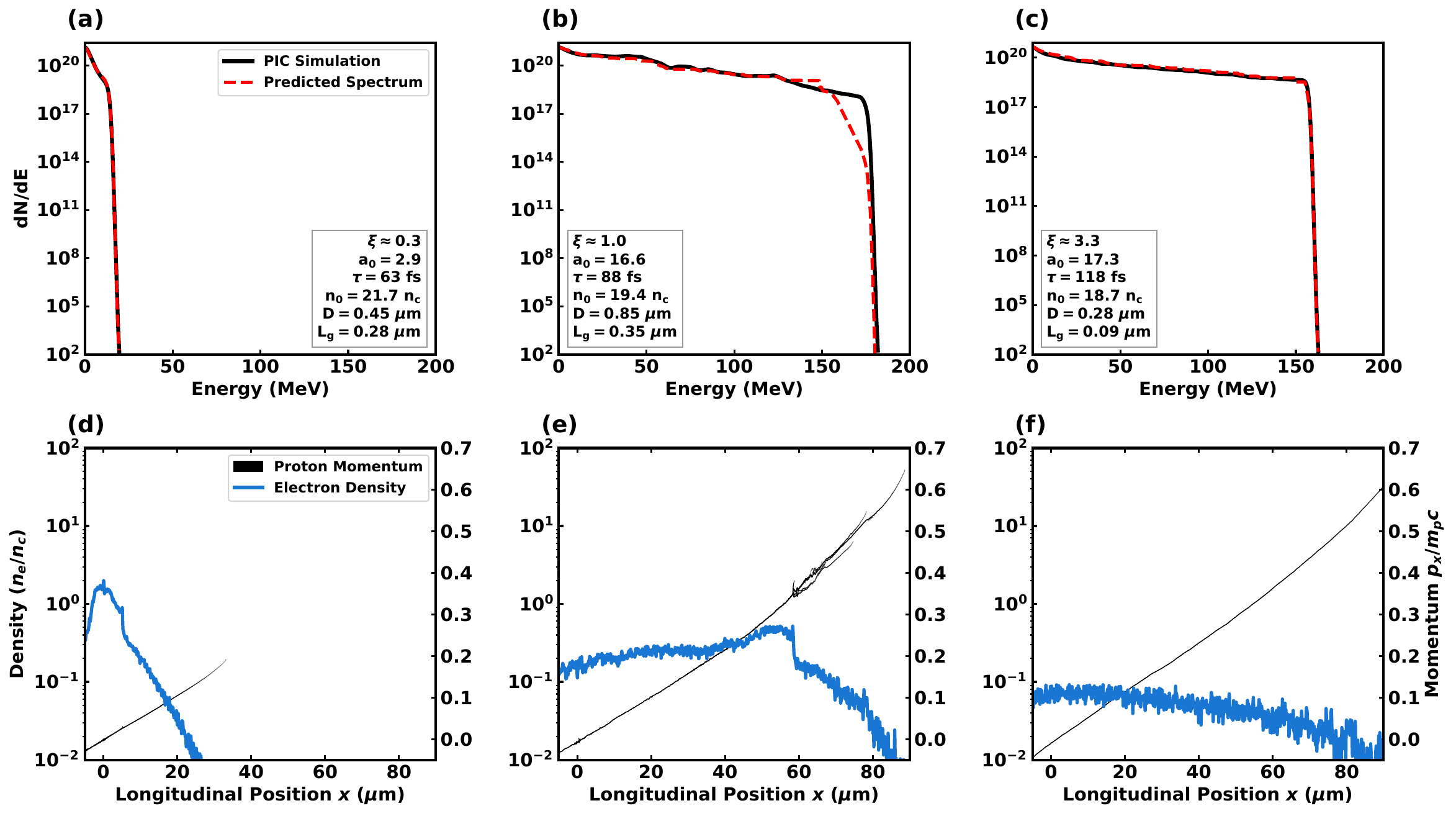}
    \caption{Kinetic validation of the surrogate-predicted mechanism transitions. The macroscopic predicted spectra (top row, a–c) are compared with the underlying microscopic plasma states (bottom row, d–f) extracted from 1D PIC simulations at $t = 800$ fs. The three cases ($\xi \approx 0.3$, $1.0$, and $3.3$) are selected as representative examples spanning the TNSA--BOA transition, rather than a controlled single-variable scan.}
    \label{fig:Fig.9}
\end{figure*}
Finally, to verify the physical validity of the surrogate's predictions, we examine whether these macroscopic transitions are supported by actual shifts in plasma kinetic dynamics. Fig.\ref{fig:Fig.9} compares the generative spectra (top row) with ground-truth PIC diagnostics. The electron density profile and the proton phase space are shown in the bottom row, which are evaluated at the asymptotic stage ($t = 800$ fs) for three representative cases.
For the opaque limit ($\xi \approx 0.3$, Fig.\ref{fig:Fig.9}(a, d)), the surrogate predicts an exponential spectral decay. The PIC diagnostics validate a surface-heating mechanism where the electron density exhibits a prominent, overdense core ($n_e > n_c$) persisting at the initial target position. The expanding proton phase space forms a mono-energetic laminar track ($\Delta p_x \approx 0$), which is the  hallmark of the quasi-static TNSA sheath.
At the surrogate-predicted RIT threshold ($\xi \approx 1.0$, Fig. \ref{fig:Fig.9}(b, e)), the kinetic state shifts. As a kinetic scar of intense volumetric heating and early-time phase-mixing, the leading edge of the proton phase space is visibly broadened, frayed, and multi-stranded. These signatures confirm the onset of BOA dynamics.
In deep transparency ($\xi \approx 3.3$, Fig. \ref{fig:Fig.9}(c, f)), the surrogate predicts an abrupt, cliff-like spectral cutoff. The corresponding PIC diagnostics reveal the electron density collapses to near-vacuum levels, and the phase-space track becomes sparse, indicating a target blowout. This analysis confirms that the surrogate is not performing blind mathematical regression. Its predicted regime boundaries reflect the real physical transitions in both macroscopic target structures and microscopic kinetic states.

\begin{figure*}[htbp]
\centering
\includegraphics[width=1.0\textwidth]{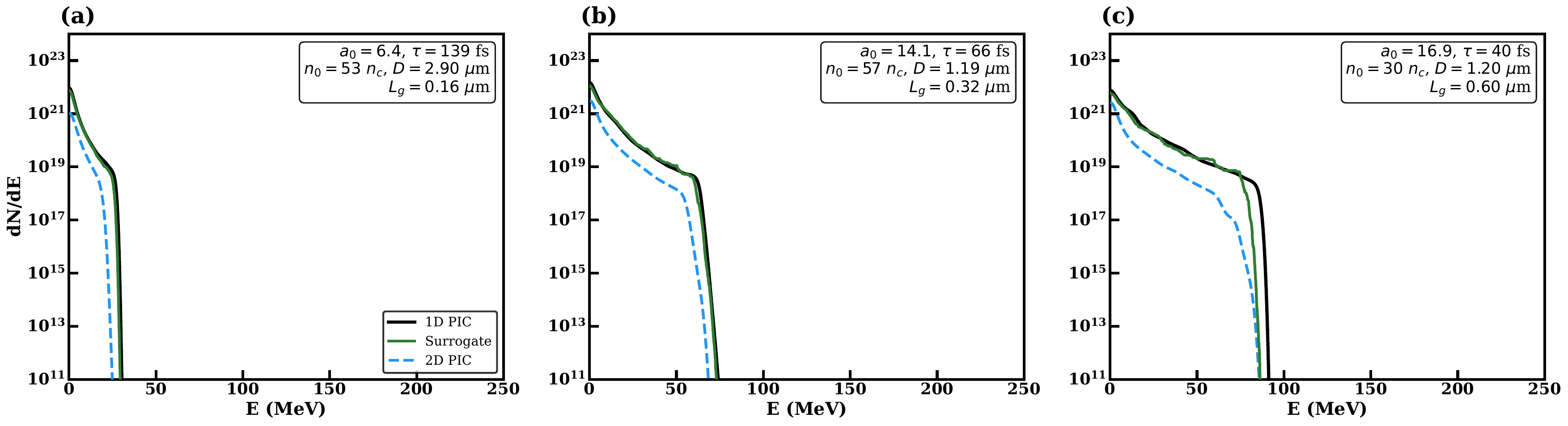}
\caption{Cross-validation of the 1D-trained surrogate against independent 2D PIC simulations for three representative cases. Black solid line: 1D PIC ground truth; green line: surrogate prediction; blue dashed line: independent 2D PIC ground truth.}
\label{fig:2dval}
\end{figure*}

To assess how the 1D-trained surrogate relates to higher-dimensional reality, we performed independent 2D PIC simulations for three representative parameter sets and compared them against both the 1D ground truth and the surrogate prediction in Fig.~\ref{fig:2dval}. To enable a consistent comparison, the 2D spectra are normalized by the transverse simulation-box width $L_y$, converting the per-unit-length spectral density to a per-unit-area measure compatible with the 1D output. This normalization is approximate, as the 2D proton distribution is not perfectly uniform along $y$; nevertheless, it permits a meaningful comparison of both spectral morphology and absolute flux magnitude. The 2D spectra reproduce the same qualitative morphology as the 1D ground truth, so that the regime classification learned by the surrogate carries over to two dimensions. The 2D cutoff energies lie below their 1D counterparts by about 15\% on average mainly due to the suppression of transverse sheath-field dilution. The reduction correlates with the laser intensity. It is largest for the lowest-$a_0$ surface-heating case, whose rear-surface sheath field is the most susceptible to transverse spreading, whereas the higher-$a_0$ cases with more penetrating coupling show smaller relative losses. The absolute spectral flux is also reduced in 2D, most pronounced in panels b and c. Both the lower cutoff and the suppressed flux are expected signatures of transverse sheath-field dilution and lateral plasma expansion that are absent in 1D. Taken together, these results support the use of the 1D-informed surrogate as a predictor of spectral morphology and regime boundaries, while the absolute cutoff energies and yields are interpreted as upper bounds.

\section{Conclusion}\label{sec:conclusion}

In conclusion, we have developed a physics-guided, decoupled dual-branch surrogate model capable of predicting continuous proton energy spectra from laser-driven ion acceleration. By integrating a $\beta$-VAE architecture for continuous spectral feature extraction with a parallel multi-layer perceptron branch for scalar boundary enforcement, the framework maps a comprehensive initial parameter space, augmented by physics-guided features, to the full spectral morphology. This allows the surrogate to achieve a predictive accuracy of $R^2=0.94$ for both $E_{\max}$ and total particle flux, a median per-sample spectral $R^2 = 0.985$ and a spectral-shape Wasserstein distance of 0.24~MeV, outperforming direct MLP regression (EMD $= 2.73$~MeV) as well as classical analytical formulations such as the Wilks scaling law.

The integration of uncertainty quantification (UQ) via deep ensembles and post-hoc variance recalibration enables the model to serve as a quantitative probabilistic diagnostic tool, demonstrating calibration errors of 6.1\% for spectra, 2.9\% for $E_{\max}$, and 3.0\% for total flux. Beyond scalar predictions, the surrogate reproduces the morphological evolution of the particle flux across distinct acceleration regimes within the 1D longitudinal framework. The data-driven predictions map signatures consistent with the critical transitions from Target Normal Sheath Acceleration (TNSA) to the Relativistically Induced Transparency (RIT) and Breakout Afterburner (BOA) regimes, as validated against the microscopic kinetic states from 1D-PIC diagnostics.

While the current 1D PIC dataset captures the dominant longitudinal electron heating and the primary transitions between TNSA and RIT/BOA regimes, this reduced dimensionality omits transverse dynamics such as Rayleigh--Taylor-like instabilities and lateral sheath expansion. In realistic multi-dimensional scenarios, these transverse effects introduce spatial dilution of the sheath field. The absolute energy and flux values predicted herein should accordingly be interpreted as upper bounds. 
In our 2D simulation, both the cutoff energy and the absolute flux are reduced.  Nevertheless, the spectral morphologies, regime-transition boundaries, and relative parametric trends remain qualitatively robust, as these features are governed by the longitudinal energy coupling dynamics that are well represented in 1D. Building on the multi-fidelity strategy of \cite{djordjevic2023transfer}, future work will leverage transfer learning to fine-tune this 1D-informed latent space using sparser multi-dimensional simulations or experimental data, thereby accommodating transverse effects. Ultimately, this physics-guided surrogate serves as a computational engine within such a multi-fidelity framework, bridging the gap between theoretical mechanism transitions and experimental optimization at advanced high-power laser facilities.
\section*{Acknowledgments}
This work is supported by the National Natural Science Foundation of China (NSFC) under Grants No.12375240, No.12535015 and the Program of China Scholarship Council (Grant CSC202506040219).

\section*{Conflict of Interest}
The authors declare no conflicts of interest.

\section*{DATA AVAILABILITY}
The datasets analyzed during the current study are available from the corresponding author on reasonable request.

\section*{Author Contributions}
\textbf{Cheng-Qi Zhang:} Conceptualization (lead); Methodology (lead); Software (lead); Investigation (lead); Formal analysis (lead); Data curation (lead); Writing original draft (lead); Writing review \& editing (lead). \\
\textbf{Yang He:} Conceptualization (supporting); Methodology (supporting); Writing review \& editing (supporting). \\
\textbf{Mamat Ali Bake:} Formal analysis (supporting); Writing review \& editing (supporting). \\
\textbf{Xilin-Wang:} Conceptualization (supporting); Methodology (supporting).\\
\textbf{Bai-Song Xie:} Supervision (lead); Funding acquisition (lead); 
Writing -- review \& editing (supporting).

\bibliographystyle{unsrt}
\bibliography{references}

\end{document}